\documentclass[aps,prl,showpacs,twocolumn,superscriptaddress]{revtex4}
\usepackage{amsmath}
\usepackage{amssymb}
\usepackage{epsfig}
\usepackage{graphicx,amsmath}
\begin{document}

\title{Magnetic-field-induced reentrance of Fermi-liquid behavior and
spin-lattice relaxation rates in ${\rm {YbCu_{5-x}Au_x}}$}

\author{V.R. Shaginyan}\email{vrshag@thd.pnpi.spb.ru}
\affiliation{Petersburg Nuclear Physics Institute, RAS, Gatchina,
188300, Russia} \affiliation{CTSPS, Clark Atlanta University,
Atlanta, Georgia 30314, USA}
\author{A.Z. Msezane}
\affiliation{CTSPS, Clark Atlanta University, Atlanta, Georgia
30314, USA}
\author{K.G. Popov}
\affiliation{Komi Science Center, Ural Division, RAS, 3a, Chernova
street Syktyvkar, 167982, Russia}
\author{V.A. Stephanovich}
\affiliation{Opole University, Institute of Mathematics and
Informatics, Opole, 45-052, Poland}
\begin{abstract}
A strong departure from Landau-Fermi liquid (LFL) behavior have
been recently revealed in observed anomalies in both the magnetic
susceptibility $\chi$ and the muon and $\rm ^{63}Cu$ nuclear
spin-lattice relaxation rates $1/T_1$ of ${\rm {YbCu_{5-x}Au_x}}$
($x=0.6$). We show that the above anomalies along with
magnetic-field-induced reentrance of LFL properties are indeed
determined by the scaling behavior of the quasiparticle effective
mass. We obtain the scaling behavior theoretically utilizing our
approach based on fermion condensation quantum phase transition
(FCQPT) notion. Our theoretical analysis of experimental data on
the base of FCQPT approach permits not only to explain above two
experimental facts in a unified manner, but to clarify the physical
reasons for a scaling behavior of the longitudinal
magnetoresistance in $\rm YbRh_2Si_2$.
\end{abstract}
 \maketitle

Landau Fermi liquid (LFL) theory designed to describe  the
thermodynamic, transport and relaxation properties of itinerant
electron systems is perhaps the most fruitful theory in condensed
matter physics \cite{land}.  The discovery of strongly correlated
states characterized by the non-Fermi liquid (NFL) behavior of
condensed matter in past century is still opening up new vistas in
physics, projecting one of the tremendous challenges in modern
condensed matter physics \cite{col,lohneysen,si}. This behavior is
so unusual that the traditional Landau quasiparticles paradigm does
not apply to it. The paradigm states that the properties is
determined by quasiparticles whose dispersion is characterized by
the effective mass $M^*$ which is independent of temperature $T$,
density $x$, magnetic field $B$ and other external parameters.

The experimental results collected on HF metals and 2D $^3$He
demonstrate the existence of very high values of a quasiparticle
effective mass $M^*$ or even its divergence \cite{lohneysen,si}.
Earlier \cite{khs,ams}, a concept of fermion condensation quantum
phase transition (FCQPT) preserving quasiparticles and intimately
related to the unlimited growth of $M^*$, had been suggested.
Further studies \cite{obz1,volovik,obz} show that it is capable to
deliver an adequate theoretical explanation of vast majority of
experimental results in different strongly correlated
Fermi-systems. In FCQPT approach, $M^*$ starts to depend on $T$,
$x$, $B$ and other external parameters. However, the extended
Landau quasiparticles paradigm is preserved. The main point here
(see, e.g., \cite{obz} and references therein) is that, as before,
the quasiparticles determine the physical properties of strongly
correlated Fermi-systems while their effective mass is a function
of external parameters. The FCQPT approach had been already
successfully applied to describe the thermodynamic properties of
such different strongly correlated systems as $^3$He on one side
and complicated heavy-fermion (HF) compounds on the other side
\cite{ckz,shag3,khodb}.

\begin{figure}[!ht]
\begin{center}
\vspace*{-0.5cm}
\includegraphics [width=0.48\textwidth]{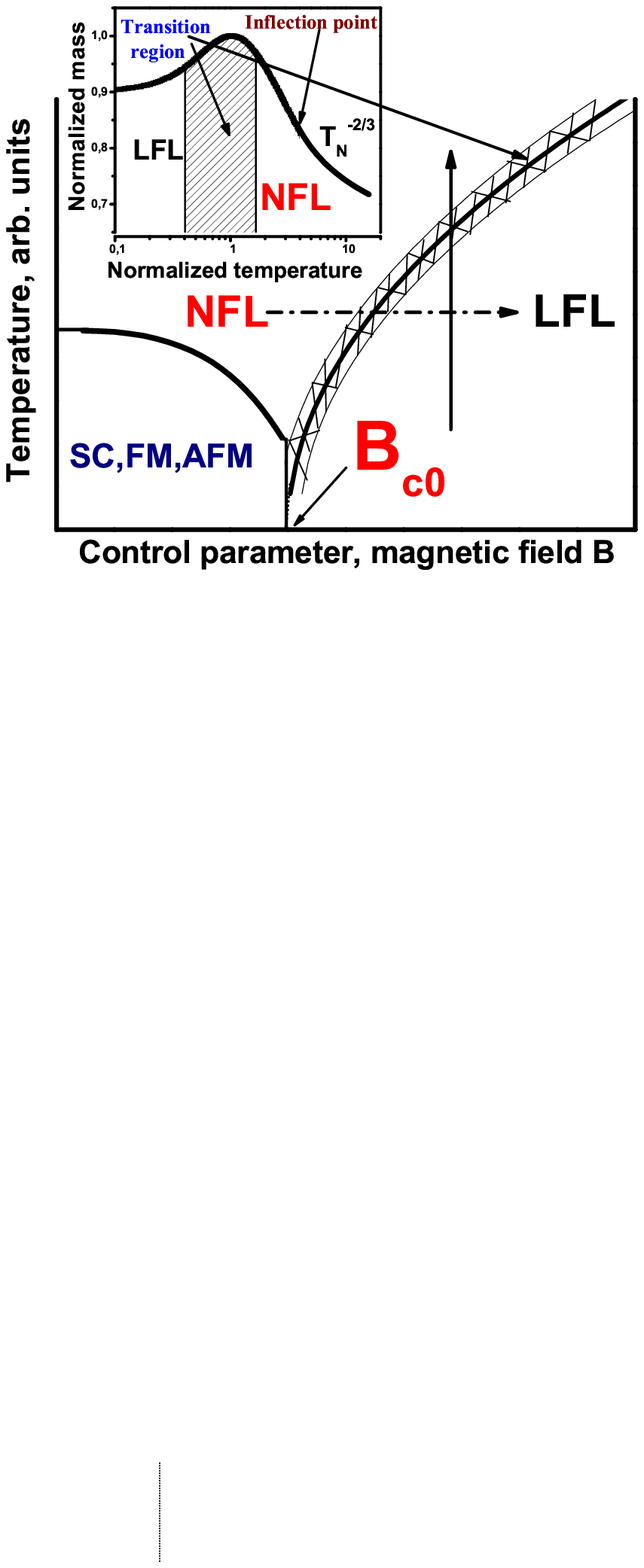}
\vspace*{-12.5cm}
\end{center}
\caption{Schematic phase diagram of HF metal. $B_{c0}$ denotes the
magnetic field at which the effective mass divergences. The
vertical arrow shows the LFL-NFL transition at fixed $B$ with $M^*$
depending on temperature. The dash-dot horizontal arrow illustrates
the system moving in the NFL-LFL direction along $B$ at fixed
temperature. At $B<B_{c0}$ the system can be in a superconducting
(SC), ferromagnetic (FM) or antiferromagnetic (AFM) states. Inset
reports a schematic plot of the normalized effective mass $M^*_N$
versus the normalized temperature. Transition regime, where $M^*_N$
reaches its maximum, is shown by the hatched area both in the inset
and in the main panel. The inflection point in $M^*_N$ is shown by
the arrow.}\label{fig:ffd}
\end{figure}

One of the most interesting and puzzling issues in the research of
HF metals is their anomalous dynamic and relaxation properties. It
is important to verify whether quasiparticles with effective mass
$M^*$ still exist and determine the physical properties of muon and
$\rm ^{63}Cu$ nuclear spin-lattice relaxation rates $1/T_1T$ in HF
metals throughout their temperature - magnetic field phase diagram,
see Fig.\ref{fig:ffd}. This phase diagram comprises both LFL and NFL
regions as well as NFL-LFL transition one (below we call it
crossover region), where magnetic-field-induced LFL reentrance
occurs. Measurements of the muon and $^{63}$Cu nuclear spin-lattice
relaxation rates $1/T_1$ in ${\rm {YbCu_{4.4}Au_{0.6}}}$ \cite{osn}
have shown that it differs substantially from ordinary Fermi liquids
obeying Korringa law. Namely, it was reported that for $T\to 0$
reciprocal relaxation time diverges as $1/T_1T\propto T^{-4/3}$
following the behavior predicted by the self-consistent
renormalization (SCR) theory. The static uniform susceptibility
$\chi$ diverges as $\chi\propto T^{-2/3}$ so that $1/T_1T$ scales
with $\chi^2$. Latter result is at variance with SCR theory
\cite{osn}. Moreover, the application of magnetic field $B$ restores
LFL behavior from initial NFL one, significantly reducing $1/T_1$
\cite{osn}. These experimental findings are hard to explain within
both conventional LFL approach and in terms of other approaches like
SCR theory \cite{kn,osn}.

In this paper, we analyze $1/T_1T$ of ${\rm {YbCu_{4.4}Au_{0.6}}}$
and show that the observed data can be well captured utilizing the
above FCQPT concept based on the extended quasiparticles paradigm.
We demonstrate that the crossover is regulated by the universal
behavior of the effective mass $M^*(B,T)$ observed in many HF
metals. It is exhibited by $M^*(B,T)$ when HF metal transits from
LFL regime (induced by a magnetic field application) to NFL one
taking place at rising temperatures. We show that violations of the
Korringa law come from the dependence of $M^*$ on magnetic field and
temperature. Our calculations of $1/T_1T$ are in good agreement with
experimental findings.

To discuss the deviations from Korringa law in light of NFL
properties of ${\rm {YbCu_{4.4}Au_{0.6}}}$, we notice that in LFL
theory spin-lattice relaxation rate $1/T_1$ is determined by the
quasiparticles near Fermi level. The above relaxation rate is
related to the decay amplitude of the quasiparticles, which in turn
is proportional to the density of states at the Fermi level
$N(E_F)$. Formally, spin-lattice relaxation rate is determined by
the imaginary part $\chi''$ of the low-frequency dynamical magnetic
susceptibility $\chi({\bf q}, \omega \to 0)$, averaged over
momentum ${\bf q}$ \cite{kn}
\begin{equation}\label{chi1}
\frac{1}{T_1}=\frac{3T}{4\mu_B^2}\sum _{\bf q}A_{\bf q}A_{-{\bf
q}}\frac{\chi''({\bf q},\omega)}{\omega},
\end{equation}
where $A_{\bf q}$ is the hyperfine coupling constant of the muon
(or nuclei) with the spin excitations at wave vector $\bf q$,
$\mu_B$ is Bohr magneton. If $A_{\bf q}\equiv A$ is independent of
$q$, then standard LFL theory relation yields
\begin{equation}\label{chi3}
\frac{1}{T_1T}=\pi A^2N^2(E_F),
\end{equation}
Equation \eqref{chi3} can be viewed as Korringa law. Since in our
FCQPT approach the physical properties of the system under
consideration are determined by the effective mass $M^*(T,B,x)$, we
express $1/T_1T$ in Eq. \eqref{chi3} via it. This is accomplished
with the standard expression \cite{land} $N(E_F)=M^*p_F/\pi^2$,
rendering Eq. \eqref{chi3} to the form
\begin{equation}\label{chi5}
\frac{1}{T_1T}=\frac{A^2p_F^2}{\pi^3}M^{*2}\equiv\eta
\left[M^*(T,B,x)\right]^2,
\end{equation}
where $\eta=(A^2p_F^2)/\pi^3=$const. The experimentally observed
relation in ${\rm {YbCu_{5-x}Au_x}}$ \cite{osn}
\begin{equation}\label{chi5a}
\frac{1}{T_1T}\propto \chi^2(T)
\end{equation}
follows explicitly from Eq. \eqref{chi5} and well-known LFL
relations $M^*\propto \chi \propto C/T$.

Having derived explicit relation between $1/T_1T$ and quasiparticle
effective mass, we are going to analyze the properties of latter.
For that, we use the model of homogeneous HF liquid with the
effective mass $M^*(T,B,x)$, where $x=p_F^3/3\pi^2$ is a number
density and $p_F$ is Fermi momentum \cite{land}. This homogeneity
permits to avoid complications associated with the crystalline
anisotropy of solids \cite{obz}. We begin with the case when at
$T\to 0$ the heavy-electron liquid behaves as LFL and is brought to
the LFL side of FCQPT by tuning of a control parameter like $x$. At
elevated temperatures the system transits to the NFL state. The
dependence $M^*(T,x)$ is governed by Landau equation \cite{land}
\begin{equation}
\frac{1}{M^*(T,x)}=\frac{1}{M}+\int\frac{{\bf p}_F{\bf p}}{p_F^3}F
({\bf p_F},{\bf p})\frac{\partial n({\bf
p},T,x)}{\partial{p}}\frac{d{\bf p}}{(2\pi)^3},\label{LQ}
\end{equation}
where $n({\bf p},T,x)$ is Fermi function, $F({\bf p}_F,{\bf p})$ is
Landau interaction amplitude and $M$ is a free electron mass. At
$T=0$, eq. \eqref{LQ} reads $M^*/M=1/(1-N_0F^1(p_F,p_F)/3)$
\cite{land}. Here $N_0$ is the density of states of a free electron
gas, $F^1(p_F,p_F)\equiv F^1(x)$ is the $p$-wave component of Landau
interaction amplitude $F$. When at some critical point $x=x_{FC}$,
$F^1(x)$ achieves certain threshold value, the denominator tends to
zero and the system undergoes FCQPT related to divergency of the
effective mass $M^*(x)/M=A+B/(x_{FC}-x)$, where $A$ and $B$ are
parameters.

A qualitative consideration of Eq. \eqref{LQ} \cite{ckz,obz1} shows
that at lowest temperatures we have the LFL regime, see the inset
to Fig. \ref{fig:ffd}. Then the system enters the transition
regime: $M^*$ grows, reaching its maximum $M^*_M$ at $T=T_M$, with
subsequent diminishing. Near temperatures $T\geq T_M$ the last
"traces" of LFL regime disappear and the NFL state takes place,
manifesting itself in decreasing of $M^*$ as $T^{-2/3}$ \cite{ckz}.
When the system is near FCQPT, it turns out that $M^*(T,x)$ can be
well approximated by a simple universal interpolating function
\cite{obz,ckz}. The interpolation occurs between the LFL
($M^*\propto T^2$) and NFL ($M^*\propto T^{-2/3}$) regimes thus
describing the above crossover \cite{ckz,obz1}. Introducing the
dimensionless variable $y=T_N=T/T_M$, we obtain the desired
expression
\begin{equation}
\frac{M^*(T/T_M)}{M^*_M} = {M^*_N(y)}\approx
c_3\frac{1+c_1y^2}{1+c_2y^{8/3}}. \label{UN2}
\end{equation}
Here $M^*_N(y)$ is the normalized effective mass,  $c_1$ and $c_2$
are parameters, obtained from the condition of best fit to
experiment, $c_3$ ensures the normalization: $M_N(1)=1$. As it
follows from Eq. \eqref{UN2}, $M^*$ reaches the maximum $M^*_M$ at
some temperature $T_M$. Since there is no external physical scales
near FCQPT point, the normalization of both $M^*$ and $T$ by
internal parameters $M^*_M$ and $T_M$ immediately reveals the
scaling behavior of the effective mass. The decay law $M^*_N\propto
T_N^{-2/3}$ along with expression \eqref{chi5} permits to express
the relaxation rate in this temperature range as
\begin{equation}\label{t43}
\frac{1}{T_1T}=a_1+a_2T^{-4/3}\propto \chi^2(T),
\end{equation}
where $a_{1}$ and $a_{2}$ are fitting parameters.

\begin{figure} [! ht]
\begin{center}
\vspace*{-0.8cm}
\includegraphics [width=0.49\textwidth]{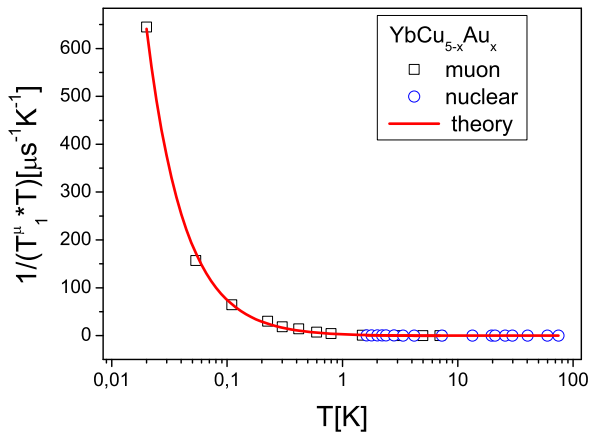}
\end{center}
\vspace*{-0.8cm} \caption{Temperature dependence of muon (squares)
and nuclear (circles) spin-lattice relaxation rates (divided by
temperature) for ${\rm {YbCu_{4.4}Au_{0.6}}}$ \cite{osn}. The solid
curve is our theoretical expression \eqref{t43}. }\label{fi3}
\end{figure}

The dependence \eqref{t43} is reported in Fig. \ref{fi3} along with experimental
points for muon and nuclear spin-lattice relaxation rates in ${\rm
{YbCu_{4.4}Au_{0.6}}}$ \cite{osn}. It is seen from Fig. \ref{fi3}
that Eq. \eqref{t43} gives pretty good description of the experiment
in the extremely wide temperature range. This means that the
extended Landau paradigm is valid and quasiparticles survive in close
vicinity of FCQPT, while the observed violation of Korringa
law comes from the dependence of the effective mass on temperature.

At small magnetic fields $B$ (that means that Zeeman splitting is
small), the effective mass does not depend on spin variable and $B$
enters Eq. \eqref{LQ} as $B\mu_B/T$ making $T_M\propto B\mu_B$
\cite{ckz,obz}. The application of magnetic field restores the LFL
behavior, and at $T\leq T_M$ the effective mass depends on $B$ as
\cite{ckz,obz}
\begin{equation}\label{B32}
M^*(B)\propto (B-B_{c0})^{-2/3}.
\end{equation}
Note that in some cases $B_{c0}=0$. In our simple model $B_{c0}$ is
taken as a parameter. We conclude that under the application of
magnetic field the variable $y=T/T_M\propto T/(\mu_B(B-B_{c0})$
remains the same and the normalized effective mass is again governed
by Eq. \eqref{UN2}. Equation \eqref{UN2} is also valid when $B$ is a
variable and $T$ is a fixed parameter, and $y=\mu_B(B-B_{c0})/T_M$
can be again considered as an effective normalized temperature. We
note that the obtained results coincide with numerical calculations
\cite{obz1,obz,ckz}.

The above considerations of the effective mass dependence on
temperature and magnetic field permit to construct the schematic
phase diagram of the substance under consideration. This diagram is
depicted in Fig. \ref{fig:ffd}. We show two LFL regions separated
by NFL one. The left LFL region may also contain magnetic
long-range order or even superconductivity. The phase boundary in
this region is the transition temperature of the corresponding
phase transitions which are incited to taking place by FCQPT
\cite{obz1,obz,khodb}. The right LFL region corresponds to that
induced by magnetic field.

\begin{figure} [! ht]
\begin{center}
\vspace*{-0.5cm}
\includegraphics [width=0.47\textwidth]{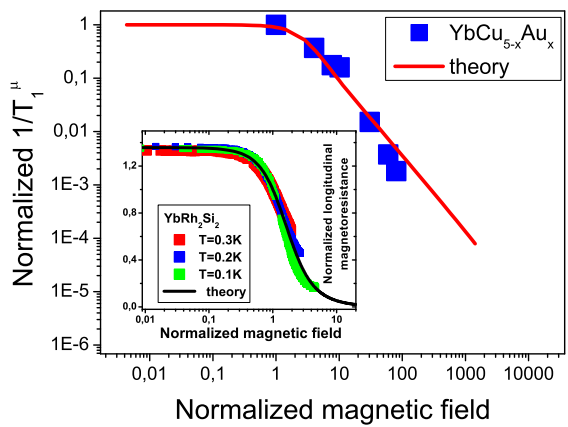}
\vspace*{-1.0cm}
\end{center}
\caption{Magnetic field dependence of normalized (in the inflection
point, see text for details) muon spin-lattice relaxation rate
$1/T_1^\mu$ in ${\rm {YbCu_{4.4}Au_{0.6}}}$ \cite{osn} along with
our theoretical B-dependence of the square of a quasiparticle
effective mass \eqref{UN2}. Inset shows the normalized
magnetoresistance $R^{\rho}_N(y)$ versus normalized magnetic field.
$R^{\rho}_N(y)$ was extracted from LMR of $\rm YbRh_2Si_2$ at
different temperatures \cite{steg} listed in the legend. The solid
line represents our calculations.}\label{fig:fi2}
\end{figure}

Figure \ref{fig:fi2} displays magnetic field dependence of
normalized (by the values of function and its argument in the
inflection point, see below) muon spin-lattice relaxation rate
$1/T_1^\mu$ in ${\rm {YbCu_{5-x}Au_x}}$ (x=0.6) \cite{osn} along
with our theoretical B-dependence. To obtain the latter theoretical
curve we (for fixed temperature and in magnetic field $B$) solve
the Landau integral equation for quasiparticle energy spectrum
$\varepsilon ({\bf k})$ (see Refs. \cite{obz1,obz} for details)
with special form of Landau interaction amplitude. Choice of the
amplitude is dictated by the fact that the system has to be in the
FCQPT point, which means that first three $k$-derivatives of the
spectrum $\varepsilon ({\bf k})$ should equal zero. Since first
derivative is proportional to the reciprocal quasiparticle
effective mass $1/M^*$, its zero (where $1/M^*=0$ and the effective
mass diverges) just signifies FCQPT, see, e.g. Refs.
\cite{obz1,obz} for details. Zeros of two subsequent derivatives
mean that the spectrum has an inflection point at Fermi momentum
$p_F$ so that the lowest term of its Taylor expansion is
proportional to $(p-p_F)^3$ \cite{ckz}. After solution of the
integral equation, the obtained spectrum had been used to calculate
an entropy $S(B,T=const)$, which, in turn, had been recalculated to
the effective mass by virtue of well-known LFL relation
$M^*(B,T)=S(B,T)/T$. We note that our calculations confirm the
validity of Eq. \eqref{UN2}. The final step was to use relation
\eqref{chi5} to calculate the reciprocal relaxation time.

The normalization procedure deserves a remark here. Namely, since
the magnetic field dependence (both theoretical and experimental)
of $1/T_1^\mu$ does not have "peculiar points" like extrema, the
normalization have been performed in the inflection point shown by
the arrow in the inset to Fig. \ref{fig:ffd}. To determine the
inflection point precisely, we first differentiate $1/T_1^\mu$ over
$B$, find the maximum of derivative and normalize the values of the
function and the argument by their values in the inflection point.
It is seen that such procedure immediately reveals the universal
magnetic field behavior of the normalized reciprocal relaxation
time $1/T_{1N}^\mu$, showing its proportionality to the effective
mass square. We emphasize here that the entire field (and
temperature) dependence of $1/T_1^\mu$ is completely determined by
corresponding dependence of the effective mass. The fact that $M^*$
becomes field, temperature and other external parameters dependent
is a key consequence of the FCQPT theory.

Consider now a longitudinal magnetoresistance (LMR)
$\rho(B,T)=\rho_0+AT^2$  as a function of $B$ at fixed $T$. In that
case, the classical contribution to LMR due to orbital motion of
carriers induced by the Lorentz force is small, while the
Kadowaki-Woods relation $K=A/\gamma_0^2\propto A/\chi^2=const$
\cite{kadw} allows us to employ $M^*$ to calculate $A\equiv A(B)$
\cite{pla3}. As a result, $\rho(B,T)-\rho_0\propto(M^*)^2$. Inset
to Fig. \ref{fig:fi2} reports the normalized magnetoresistance
\begin{equation}\label{rn}
R_N^\rho(y)=\frac{\rho(y)-\rho_0}{\rho_{inf}}\propto\frac1{T_{1N}^\mu}
\propto (M_N^*(y))^2
\end{equation}
vs normalized magnetic field $y=B/B_{inf}$ at different
temperatures, shown in the legend. Here $\rho_{inf}$ and $B_{inf}$
are LMR and magnetic field taken at the inflection point shown in
the inset to Fig. \ref{fig:ffd} by the arrow. The transition region
where LMR starts to decrease is shown in the inset by the hatched
area and takes place when the system moves along the horizontal
dash-dot arrow. We note that the same normalized effective mass has
been used to calculate both $1/T_{1N}^\mu$ and the normalized LMR.
Thus, Eq. \eqref{rn} determines the close relationship between the
quite different dynamic properties, showing the validity of the
extended Landau paradigm.

Both theoretical and experimental curves have been normalized by
their inflection points, which also reveals the universal behavior
- the curves at different temperatures merge into a single one in
terms of scaled variable $y$. Figure \ref{fig:fi2} shows clearly
that both normalized magnetoresistance $R^{\rho}_N$ and reciprocal
spin-lattice relaxation time well obeys the scaling behavior given
by Eq. \eqref{rn}. This fact obtained directly from the
experimental findings is a vivid evidence that both above
quantities behavior is predominantly governed by field and
temperature dependence of the effective mass $M^*(B,T)$.

In summary, our theoretical study of $1/T_1T$ and LMR in two
different HF compounds shows that their characteristic behavior is
due to the dependence of the quasiparticle effective mass on
magnetic field, temperature and other external parameters. Our
results are in good agreement with experimental facts and allow us
to confirm the validity of the extended Landau paradigm. This
paradigm, in turn, permits us to explain for the first time the
magnetic field behavior of both $1/T_1T$ and LMR.

This work was supported in part by the grants: RFBR No.
09-02-00056, DOE and NSF No. DMR-0705328, and Opole University
Intramural Grant (Badania Statutowe).

\end{document}